\documentclass[twocolumn,showpacs,preprintnumbers,amsmath,amssymb]{revtex4}
\usepackage{graphicx}
\usepackage{multirow}
\usepackage{color}

\begin{document}

\title{Velocity-selected molecular pulses produced by an electric guide}

\author{C. Sommer}
\author{M. Motsch}
\author{S. Chervenkov}
\author{L.D. van Buuren}
\author{M. Zeppenfeld}
\author{P.W.H. Pinkse} \altaffiliation[Present address: ]{Universiteit Twente, Mesa+ Institute for Nanotechnology, Postbus 217, 7500AE Enschede, The Netherlands.}
\author{G. Rempe}
\affiliation{Max-Planck-Institut f{\"u}r Quantenoptik, Hans-Kopfermann-Str. 1, 85748 Garching, Germany. E-mail: christian.sommer@mpq.mpg.de}

\begin{abstract}
Electrostatic velocity filtering is a technique for the production of continuous guided beams of slow polar molecules from a thermal gas.
We extended this technique to produce pulses of slow molecules with a narrow velocity distribution around a tunable velocity. The pulses are generated by sequentially switching the voltages on adjacent segments of an electric quadrupole guide synchronously with the molecules propagating at the desired velocity. This technique is demonstrated for deuterated ammonia (ND$_{3}$), delivering pulses with a velocity in the range of $20-100\,\rm{m/s}$ and a relative velocity spread of $(16\pm 2)\,\%$ at FWHM. At velocities around $60\,\rm{m/s}$, the pulses contain up to $10^6$ molecules each. The data are well reproduced by Monte-Carlo simulations, which provide useful insight into the mechanisms of velocity selection.

\end{abstract}


\maketitle


\section{Introduction}
\label{intro}
Beams of cold molecules offer exciting prospects for experiments in physics and chemistry \cite{Dulieu2006,Smith2008,Carr2009,Krems2009}.
This includes cold-collision studies and cold reaction dynamics \cite{Krems2005,Gilijamse2006,Krems2008,Willitsch2008,Sawyer2009,Zuchowski2009} as well as high-resolution experiments to determine, for example, the electric dipole moment (EDM) of the electron \cite{Hinds1997,Meyer2008,Tarbutt2009,Vutha2009}. For these measurements a beam of slow molecules with well-defined velocity and a high degree of internal-state purity is advantageous or even mandatory. Velocity-selected pulses of slow cold molecules have been produced so far only by deceleration techniques \cite{Bethlem1999,Gupta2001,Fulton2004,Nar08b,Hogan2009}. In this work we present an alternative method based on velocity filtering.

In its original concept, the technique of velocity filtering by an electric or magnetic guide \cite{Rangwala2003,Junglen2004a,Patterson2007} is a method for producing a continuous beam of slow molecules. The main advantage of this method lies in its simplicity and in the high flux obtainable at low average velocities. In addition, it is a very general method since it is applicable to polar (electric guide) and paramagnetic (magnetic guide) molecules \cite{Sommer2009,Krems2009} as long as they have a reasonably large positive Stark shift (linear or quadratic \cite{MotschWater}) or Zeeman shift. Besides, no modifications of the guide are required when changing molecular species.
Typically, these molecules are extracted from a thermal effusive source. The extracted beam therefore contains molecules populating different internal states \cite{Motsch2007}. By adding a buffer-gas cooling scheme prior to velocity filtering, the purity of the internal-state distribution can be strongly increased \cite{Maxwell2005,vanBuuren2009,Sommer2009}.

To obtain monokinetic molecular pulses, we have now extended the electrostatic guiding technique by including velocity selection. The latter is reminiscent of the experiment carried out by Eldridge \cite{Eldridge1927} in 1927 and employed in many other experiments afterwards \cite{Scoles1988}, in which velocity selection was achieved by mechanically chopping a continuous molecular beam effusing from a slit. In our setup the selection is done electrically by switching the voltages applied to the electrode segments of the guide on and off in a specific sequence. The switching times are adjusted to the desired molecular velocity. In this way molecular pulses with a narrow velocity distribution around a desired velocity are obtained. Our velocity-selection technique is highly versatile, demonstrating the generality of the electric-guiding technique. It allows for an immediate transition from a continuous guided beam to a pulsed mode of operation along with tuning to different desired velocities.

\section{Experimental Setup}
\label{exper}
\begin{figure}[t]
\begin{center}
\includegraphics[scale=0.5]{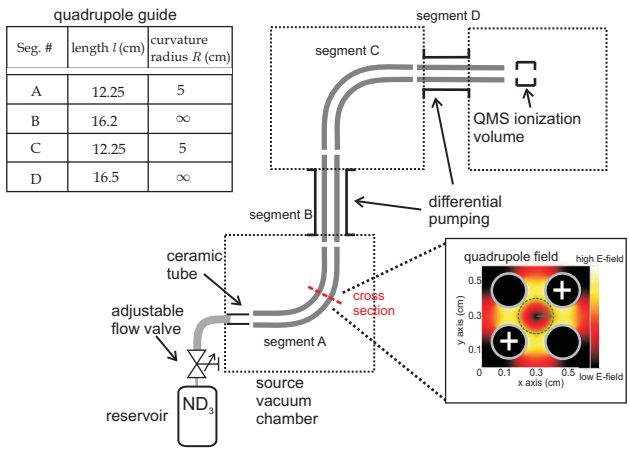}
\caption{(Color online) The experimental setup consists of three differentially-pumped vacuum chambers. In the first vacuum chamber molecules are injected into the guide through a ceramic nozzle. A fraction of these molecules is guided via the four-segment electric guide to a quadrupole mass spectrometer (QMS) where they are detected. The inset in the lower right part of the figure depicts the transverse profile of the electric field. $r$ is the free inner radius of the guide, at which the maximum trapping field is reached. The lengths, $l$, and curvatures, $R$, of the four segments are detailed in the table in the upper left corner of the figure.}
\label{fig1}
\end{center}
\end{figure}

The setup used to guide polar molecules is shown schematically in Fig.~\ref{fig1}. This is the same setup used for previous experiments and has been described in detail elsewhere \cite{MotschBoosting,MotschWater}. Briefly, it consists of three differentially pumped vacuum chambers accommodating the electrodes of the electric guide.
Four parallel stainless-steel rods with a diameter of $2\,\rm{mm}$ in a quadrupole configuration with $1\,\rm mm$ spacing between the rods constitute the electrodes of the guide. Alternating high voltages on neighboring electrodes give rise to a quadrupole field, which has a field minimum at the center (see inset, Fig.~\ref{fig1}). Thus polar molecules in low-field-seeking (lfs) states, experiencing a positive Stark shift, are transversely confined, provided their kinetic energy does not exceed the potential barrier (trap depth). The Stark shift $\Delta W^{s}(E_{\rm max})$ at the maximum of the trapping field $E_{\rm max}$ determines the maximum transverse velocity $v_{t,\,\rm max}= [v_{x}^{2} + v_{y}^{2}]^{1/2} = \sqrt{2\Delta W^{s}(E_{\rm max})/m}$. A bend in the guide's electrodes limits the molecular velocity in the longitudinal direction. The longitudinal cutoff velocity $v_{z,\,\rm max} = \sqrt{\Delta W^{s}(E_{\rm max})R/(rm)}$ is obtained by equating the centrifugal force $mv^{2}/R$ to the restoring force caused by the Stark shift $\sim \Delta W^{s}/r$, where $R$ is the radius of curvature of the guide electrodes (see Fig.~\ref{fig1}) and $r$ is the free inner radius of the guide, at which the maximum trapping field is reached (see inset of Fig.~\ref{fig1}).

The guide is split into four segments. The guide segments $A$ and $C$ made of bent electrodes are located in the first and second vacuum chamber, respectively, while the straight segments $B$ and $D$ are placed in the differential pumping sections connecting the vacuum chambers. The guide segments are separated from each other by gaps of $1\,\rm{mm}$. The molecules are injected into the guide in the first vacuum chamber through a ceramic tube with a diameter of $1.5\,\rm{mm}$ connected to a gas reservoir. The ceramic tube is detached from the guide by a gap of $1\,\rm{mm}$. The guided molecules are detected in the third vacuum chamber by a quadrupole mass spectrometer (QMS, Pfeiffer QMG422) positioned $2.2\,\rm{cm}$ downstream from the guiding electrodes. The guided molecules are ionized by electron impact and mass-filtered. Single-ion detection is achieved by employing a secondary electron multiplier.\\

\begin{figure*}[t]
\begin{center}
\includegraphics[scale=0.5]{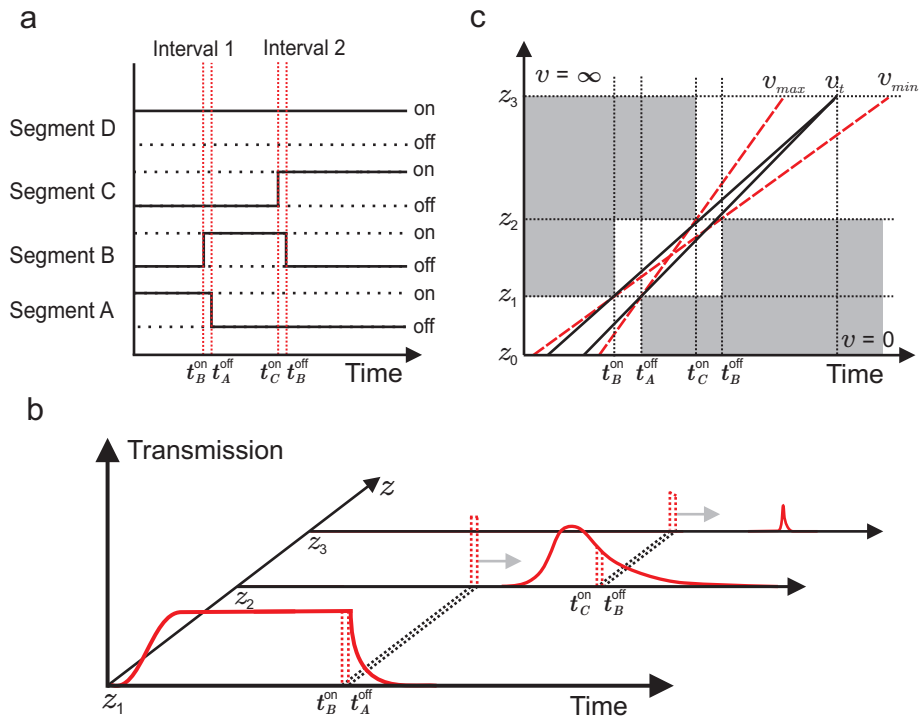}
\caption{(Color online) In (a) the switching sequence of the electrodes is presented. In the time intervals $[t_B^{\rm on}, t_A^{\rm off}]$ and $[t_C^{\rm on}, t_B^{\rm off}]$, respectively, two consecutive segments are in guiding configuration. The pulse-production process is sketched in (b). $z_{1}$, $z_{2}$, and $z_{3}$ are the distances from the exit of the ceramic tube to the entrance of segment $B$, to the entrance of segment $C$, and to the QMS, respectively. The transmission for certain velocities is schematically presented in the $z-t$ diagram in (c). The slopes of the black lines represent molecules from different starting times that can overlap each other at the QMS position. 
The slopes of the red dashed lines show the maximal and minimal velocities of the molecules in the pulse.
}
\label{fig2}
\end{center}
\end{figure*}

\section{Velocity Selection Scheme}
The molecules which are trapped in the electrostatic guiding potential span a large continuum of longitudinal velocities. 
In our experiment we start with a continuous flow of guided molecules whose flux is described by the relation \begin{eqnarray}\Phi \propto \int_0^{v_{z,\,\rm max}} v_{z}f(v_{z})dv_{z}, \label{eqn1}\end{eqnarray} where $f(v_{z})$ is the velocity distribution of the molecules with respect to their longitudinal velocity $v_z$. The integration is performed over the interval of longitudinal velocities $v_{z}$ from 0 to the maximum longitudinal velocity in the beam, $v_{z,\,\rm max}$. This implies that under steady-state conditions molecules with all possible longitudinal velocities $v_z$ are present in the guide at every instant of time. To filter only molecules with a certain longitudinal velocity out of the total flux, the velocity-dependent selection scheme described below is applied. In order to avoid modification of the velocity distribution stemming from collisions of slow molecules with fast molecules near the exit of the ceramic tube \cite{MotschBoosting}, all our experiments are performed at a low reservoir pressure ($1\times10^{-2}\,\rm{mbar}$).

As described in the previous section, the guide is composed of four segments, which can be switched independently to a guiding or non-guiding configuration. By applying an appropriate switching sequence to the guide segments ideally only molecules that continuously experience a guiding field are steered to the end of the guide. Most of the other molecules are lost from the guide and do not reach the detector. This results in molecular pulses characterized by a certain velocity and velocity spread. The process is schematically presented in Fig.~\ref{fig2}. At time $t=0$ segment $A$ is in guiding configuration while segments $B$ and $C$ are switched off. Segment $D$ is on all the time to avoid switching transients in the response of the mass spectrometer. Segment $A$ remains on for time $t_A$ ($t_A\equiv t_A^{\rm off}$), during which molecules with all possible longitudinal velocities below $v_{z,\,\rm max}$ are guided. They cannot, however, propagate farther through segment $B$. At time $t_B^{\rm on}$ ($t_B^{\rm on}< t_A^{\rm off}$), shortly before segment $A$ is turned off, segment $B$ is turned on. This gives rise to an overlap time interval $\Delta t_{AB}=t_A^{\rm off}-t_B^{\rm on}$ during which both segments are in guiding configuration. This ensures that in this time window molecules in the gap and its immediate vicinity will experience a continuous electric quadrupole field and can enter the subsequent segment without being disturbed by the switching of the electric field. Thus the overlap interval defines a molecular pulse with duration $\Delta t_{AB}$ containing molecules with all longitudinal velocities below the cutoff velocity $v_{z,\,\rm max}$. This pulse traverses the gap and is launched into segment $B$, while segment $C$ remains off. The time interval during which segment $B$ is in guiding configuration depends on the desired longitudinal velocity $v_{z,\,\rm set}$ according to $t_{B} = l_{B}/v_{z,\,\rm set}$, where $l_{B}$ is the length of segment $B$. Just before segment $B$ is switched off at time $t_B^{\rm off}$, segment $C$ is turned on at time $t_C^{\rm on}$, resulting in the overlap interval $\Delta t_{BC}=t_B^{\rm off}-t_C^{\rm on}$. This opens up a pathway for molecules to bridge the gap and enter segment $C$. Segment $A$ remains in non-guiding configuration. It is important to point out that when moving along segment $B$, the initially short molecular pulse spreads along the propagation line as a result of longitudinal velocity dispersion. Molecules that arrive at the gap between segments $B$ and $C$ before segment $C$ is switched on are lost. The same occurs for molecules which are too slow and are still in segment $B$ when the latter is turned off. Only the molecules whose velocity is matched to their arrival time at the gap between segments $B$ and $C$ experience a continuous guiding field and are accepted by segment $C$.

Since the overlap intervals $\Delta t_{AB}$ and $\Delta t_{BC}$ are not infinitesimally short, the velocity distribution of the molecules has a finite width around the selected velocity. The center velocity as well as the minimum and the maximum velocity of the molecules comprising the output molecular pulse arriving at the detector can be graphically derived with the help of the $z-t$ diagram shown in Fig.~\ref{fig2} (c). The guiding configurations are designated by the white boxes, while the grey shading designates the non-guiding configurations. The slopes of the black lines represent the lower and the upper boundaries for the velocity of the molecules arriving at the detector at time $t$. The slopes of the red dashed lines mark the lowest and the highest admissible velocities in the produced pulse.

From the switching times, the maximal repetition rate of the pulses can be determined. The rate is given by the time the molecules need to move from the entrance of the first segment to the exit of the second segment. This stems from the fact that the first segment can be switched on again only when the molecular pulse has entered the third guide segment. In our setup repetition rates of the order of a few hundred Hertz can be realized for molecular pulses with velocities of a few tens of $\rm{m/s}$. To give an example, at a velocity of $60\,\rm{m/s}$ a maximal repetition rate of $\sim 210\,\rm{Hz}$ can be employed. In principle, higher repetition rates are attainable by using a configuration with shorter segments.

\section{Results}
\label{results}
To determine the properties of the molecular pulses we have performed time-of-flight measurements. The obtained information includes the shape of the pulse, its intensity, its width, and its center velocity.
\begin{figure}[t]
\begin{center}
\includegraphics[scale=0.45]{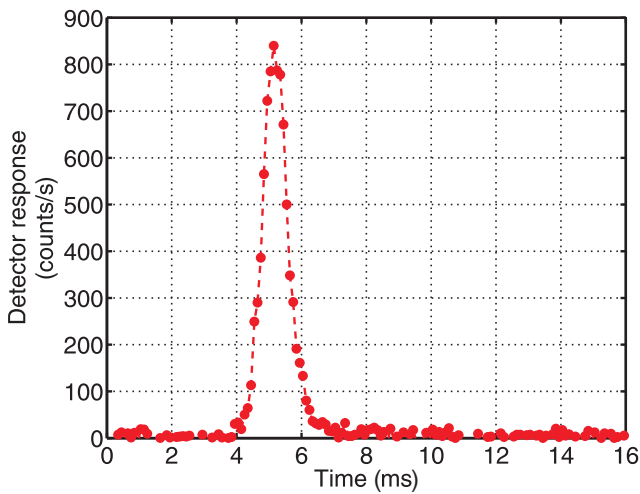}
\caption{
(Color online) Time-of-flight signal for a pulse with a center velocity of $60\,\rm{m/s}$. The detector response is plotted against the arrival time of the molecules. The middle of the time interval $[t_C^{\rm on},t_B^{\rm off}]$ corresponds to $t=0$ in the plot. The distance the molecules traverse during the time-of-flight measurement is the sum of the lengths of segments $C$ and $D$, and the gap of $2.2\,\rm{cm}$ between the end of the guide and the QMS. The depicted pulse has been obtained with the gradient off-configuration (see text). Background contributions have been subtracted from the raw data.}
\label{fig3}
\end{center}
\end{figure}
Fig.~\ref{fig3} shows the time-of-flight signal $S(t)$ of a molecular pulse with a center velocity of $60\,\rm{m/s}$. The velocities of the detected molecules are determined from their arrival times by converting the acquired signal $S(t)$ from the time domain to the velocity domain. This is done via the relation $S(t) \Delta t = [S(t) \times (t^{2}/l)]\Delta v_z = P(v_z) \Delta v_z$, where $t$ stands for the arrival time of the molecules relative to the middle of the time interval $[t_C^{\rm on},t_B^{\rm off}]$, and $l$ is the traversed length. The bin width of the histogram, $\Delta t$, is much smaller than all relevant time scales.

\begin{figure}[t]
\begin{center}
\includegraphics[scale=0.45]{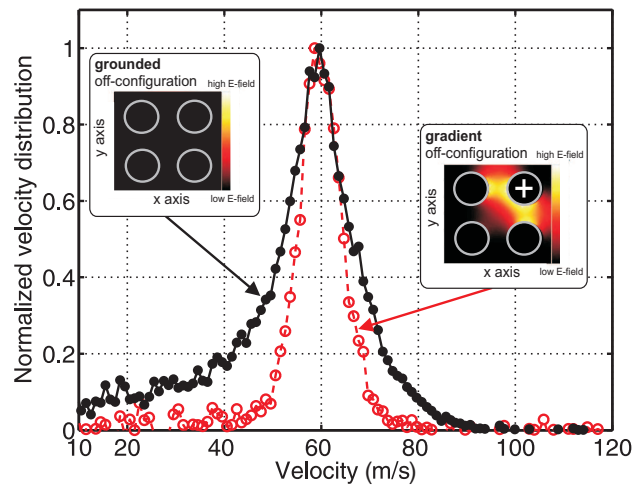}
\caption{(Color online) Normalized velocity distributions of pulses with an average velocity of $60\,\rm{m/s}$ measured in two modes of operation: with the grounded off-configuration (black filled dots) and with the gradient off-configuration (red open dots). For the guiding configuration, $+4\,\rm{kV}$ are applied to one pair of opposite electrodes while the other pair of opposite electrodes is at $0\,\rm{V}$, this leading to electric fields of up to $\sim 40\,\rm{kV/cm}$. In the gradient off-configuration the deflection field reduces the amount of molecules that can still enter the successive guide segment. The asymmetric shape of the grounded off-configuration peak results from the bend geometry of segment $C$ of the guide.}
\label{fig4}
\end{center}
\end{figure}

To rationalize the pulse formation we analyze velocity distributions at a selected center velocity of $60\,\rm{m/s}$ (see Fig.~\ref{fig4}). First, we performed time-of-flight measurements switching off all four electrodes of the respective segments to achieve the non-guiding regime. This non-guiding configuration of the electrodes is referred to as grounded off-configuration (Fig.~\ref{fig4} (left inset)). The resulting velocity distribution is given by the black filled-dotted curve in Fig.~\ref{fig4}. The broad shape results from molecules that survive in the guide during the off-configuration. In our further discussion these molecules are referred to as residual molecules. These molecules are not guided by an electric field but manage to reach the next guiding segment simply by free rectilinear flight. The asymmetric shape of the broader velocity distribution results from the bend of segment $C$ after the gap (see Fig.~\ref{fig1}). Residual molecules whose velocities are higher than the desired velocity $v_{z,\,\rm set}$ set by the switching time $t_C^{\rm on}$ can make it through the gap and enter segment C by free rectilinear flight. If these molecules reach the bend of segment $C$ at times $t<t_C^{\rm on}$ they escape from the guide. This leads to a reduction of the broadening at the high-velocity side of the velocity distribution. On the other hand, residual molecules whose velocities are smaller than the desired velocity $v_{z,\,\rm set}$, can remain in segment $B$ and manage to reach segment $C$ even at times $t>t_B^{\rm off}$ by free rectilinear flight. These molecules are guided in segment $C$ since this segment is on for times $t>t_B^{\rm off}$, and thus contribute to the broadening of the velocity distribution at its low-velocity side.
To reduce the contribution of residual molecules to the pulse, we applied another non-guiding configuration, termed gradient off-configuration, in which three of the electrodes of the guide are switched off and one of the electrodes remains at $+4\,\rm{kV}$ (Fig.~\ref{fig4} (right inset)). This non-guiding scheme results in a deflection field of $13\,\rm{kV/cm}$ in the center of the guide pushing the residual molecules out of the guide.  This leads to the narrowed velocity distribution given by the red open dots in Fig.~\ref{fig4}. The temporal profile of the corresponding pulse is shown in Fig.~\ref{fig3}.

The overlap intervals have been optimized to maximize the number of molecules per pulse without significantly increasing the width of the velocity distribution. The duration of the overlap intervals has been adjusted proportional to $1/v_{z,\,\rm set}$ to ensure that for every $v_{z,\,\rm set}$ the molecules travel the same distance $d$ during the overlap times. In our experiment $d=8.4\,{\rm mm}$. In this way molecules of all velocities experience the same switching transient fields. For our setup optimal overlap times around hundred microseconds are found. For example, an overlap time of $140\,\rm{\mu s}$ has been used for $\Delta t_{AB}$ and $\Delta t_{BC}$ for the data shown in Fig.~\ref{fig4}.

An important parameter of the segmented-guiding technique is the number of molecules in a pulse. To determine this value we have used the background-subtracted data of the histograms representing single-molecule detections of the QMS. Using a previous calibration of the QMS \cite{Sommer2009}, we have estimated that the pulse at $60\,\rm{m/s}$ contains $10^5$ molecules. This number can be increased to $10^{6}$ molecules by increasing the pressure in the reservoir \cite{MotschBoosting}.

\begin{figure}[t]
\begin{center}
\includegraphics[scale=0.45]{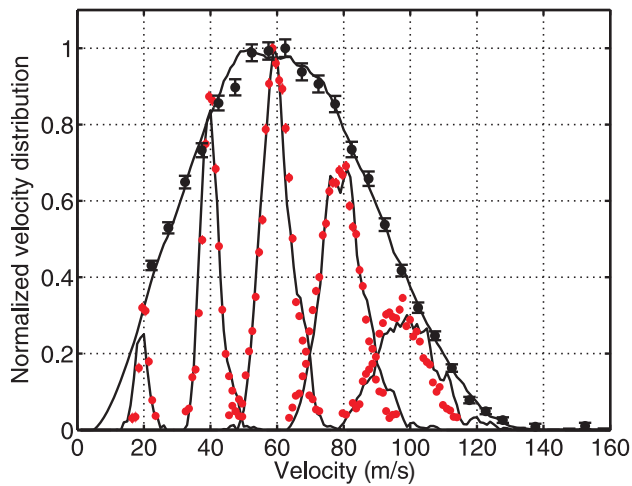}
\caption{(Color online) Experimental velocity distributions obtained from the continuous beam (black dots with error bars from statistical uncertainties) and from the segmented guiding (red dots). Monte-Carlo simulations of the velocity distributions are shown for continuous and pulsed operation of the molecular guide by black solid curves. Both experimental and simulated velocity distributions have been rebinned to $1\,\rm{m/s}$.
}
\label{fig5}
\end{center}
\end{figure}

As a next step, we compare the velocity distributions at different center velocities. In Fig.~\ref{fig5} the experimentally determined velocity distributions of molecular pulses for several center velocities are displayed by red dots. Overall, good agreement between the heights of the velocity distributions of the velocity-selected pulses and the measured velocity distribution for a continuous beam of guided molecules is found. This demonstrates that electric switching does not lead to a reduction of the number of molecules at the desired center velocity of the pulse. We have also successfully extracted velocity-selected molecular pulses from our buffer-gas cooled source \cite{vanBuuren2009,Sommer2009} to demonstrate that the same technique can be employed to create velocity-selected pulses with a high degree of internal-state purity.

To further substantiate the experimental results we compare them to the results from Monte-Carlo simulations for a bent electric guide. The simulations employed a model guide accounting for all physical aspects of the experiment and propagated molecules with the appropriate parameters and starting conditions. The input range of longitudinal velocities was from $10\,\rm m/s$ to $240\,\rm m/s$. Simulations for pulsed operation as well as for continuous guiding (with all segments of the guide on at all times) have been carried out. As in the experiment, the velocities are obtained from their arrival time in the detector with respect to the time $t_C^{\rm on}+\frac{1}{2} \Delta t_{BC}$. Due to the relatively large statistical uncertainties the simulated peaks have been smoothed using the Savitzky$-$Golay smoothing filter \cite{Press1996} employing a third-order polynomial with 6 points. The simulated continuous velocity profile has been smoothed as well using the above method with 30 points. The results of the simulations are presented in Fig.~\ref{fig5} (black solid curves). Good agreement is found between the experimental data and the simulations for the continuous beam as well as for the pulses at different longitudinal velocities $v_{z,\,\rm set}$.

The full widths at half maximum (FWHM) of the velocity distributions, $\Delta v_z$, are obtained from fits with Voigt profiles \cite{Olivero1977}. In this way the ratios $\Delta v_{z,\,\rm set}/v_{z,\,\rm set}$ can be derived. Overall, a value of $\Delta v_{z,\,\rm set}/v_{z,\,\rm set} = (16\pm 2)\,\%$ is obtained for the relative velocity spread. The non-zero width of the velocity distributions is attributed to two effects, the finite duration of the overlap intervals proportional to $1/v_{z,\,\rm set}$ and the presence of residual molecules. In the following we describe these two effects and their contributions to the widths of the velocity distributions in more detail.

The desired longitudinal velocity is determined by $v_{z,\,\rm set} = l_B/t_B$ (see Section 3). By differentiating this expression with respect to time and by substituting $l_B/v_{z,\,\rm set}$ for $t_B$, we obtain the following relationship between the pulse width $\Delta t_B$ and the width of its velocity distribution, $\left| \Delta v_{z,\,\rm overlap}\right | = (v_{z,\,\rm set}^2/l_B)\Delta t_B$. The two short but non-zero overlap intervals $\Delta t_{AB}=\Delta t_{BC}=d/v_{z,\,\rm set}$ ($d<<l_{\rm B}$) determine a triangular velocity distribution resulting from a convolution of two rectangular velocity distributions. Therefore we obtain the following relation for the FWHM of the velocity distribution stemming from the finite overlap intervals,

\begin{eqnarray}
\Delta v_{z,\,\rm overlap} = \frac{d}{l_B} v_{z,\,\rm set}. \label{eqn2}
\end{eqnarray}

We determine that $\Delta v_{z,\,\rm overlap}/v_{z,\,\rm set}=5\%$ for our settings. From the measured widths we can conclude that residual molecules substantially contribute to the observed broadening. Therefore, decreasing the overlap intervals to smaller values does not lead to much narrower velocity distributions. Indeed, we have observed that for shorter overlap intervals of a few tens of microseconds the widths of the velocity distributions do not decrease significantly. They are an order of magnitude broader than expected from Eq.~\ref{eqn2} for these short overlap times. Additionally, a substantial reduction in peak height is observed for the slowest pulse with a center velocity of $20\,\rm{m/s}$. This additional reduction has not been seen with overlap times of a few hundred microseconds.

The residual molecules manage to reach the subsequent segment shortly before or shortly after the overlap intervals. The number of these molecules depends on the mean survival time $\tau$ they can spend in gradient off-configuration before being kicked out by the deflection field. This time $\tau$ depends on the transverse velocity of the molecules and on the strength of the deflection field. For a straight guide, both the transverse velocity and the strength of the deflection field are independent of the set velocity $v_{z,\,\rm set}$. Indeed, we have verified this by fitting the transverse velocity distributions obtained from the Monte-Carlo simulations with Gaussian profiles. The fits demonstrate that for all longitudinal velocities $v_{z,\,\rm set}$ the average value of the transverse velocity roughly equals $7\,\rm m/s$.

To account for the broadening resulting from residual molecules in addition to the broadening stemming from the finite overlap time, we use the following model including both effects. Let $t_B=t_B^{\rm off}-t_B^{\rm on}$ (see Fig.~\ref{fig2}). Let us also first assume that there are no residual molecules in the vicinity of the gap between segments $A$ and $B$, and the overlap time $\Delta t_{AB}=0$. This assumption implies that the broadening of the velocity distribution originates only from the finite overlap time $\Delta t_{BC}$ and from residual molecules in the vicinity of the gap between segments $B$ and $C$. The minimum velocity $v_z^{\rm min}$ at which molecules in segment $B$ can still reach the subsequent segment $C$ is $v_z^{\rm min} = l_B/[t_B+\tau]$. The maximum velocity $v_z^{\rm max}$ at which some molecules remain guided in segment $C$ is $v_z^{\rm max} = l_B/[t_B-\tau-\Delta t_{BC}]$. After substituting $t_B=l_B/v_{z,\,\rm set}$, the above expressions are transformed into $v_z^{\rm min}=l_B v_{z,\,\rm set}/[l_B+v_{z,\,\rm set}\tau]$ and $v_z^{\rm max}=l_B v_{z,\,\rm set}/[l_B-v_{z,\,\rm set}(\tau+\Delta t_{BC})]$, respectively. The difference between $v_z^{\rm max}$ and $v_z^{\rm min}$ defines a rectangular distribution of longitudinal velocities determined by the overlap time $\Delta t_{BC}$ and the survival time $\tau$ of the residual molecules in the vicinity of the gap between segments $B$ and $C$. By analogy, the same model is applicable to the gap between segments $A$ and $B$ if we assume that now there are residual molecules only in the vicinity of the gap between segments $A$ and $B$, $\Delta t_{AB}>0$, and $\Delta t_{BC}= 0$. The overall velocity distribution is thus obtained by the convolution of the two rectangular velocity distributions. The resulting velocity distribution has a triangular profile with a FWHM given by the formula
\begin{eqnarray}
\Delta v_{z} =\frac{l_B v^2_{z,\,\rm set} (2 \tau + \Delta t_{BC})}{(l_B - v_{z,\,\rm set} \tau)(l_B + v_{z,\,\rm set}(\tau + \Delta t_{BC}))} \label{eqn3}
\end{eqnarray}
for a guide without bends. It is obvious that for infinitesimally short survival times $\tau$ ($\tau \rightarrow 0$) and for overlap times satisfying the condition $d<<l_{B}$, the above formula reduces to Eq.~\ref{eqn2}. In this model the survival time $\tau$ can be considered as an effective additional overlap time.

\begin{figure}[t]
\begin{center}
\includegraphics[scale=0.45]{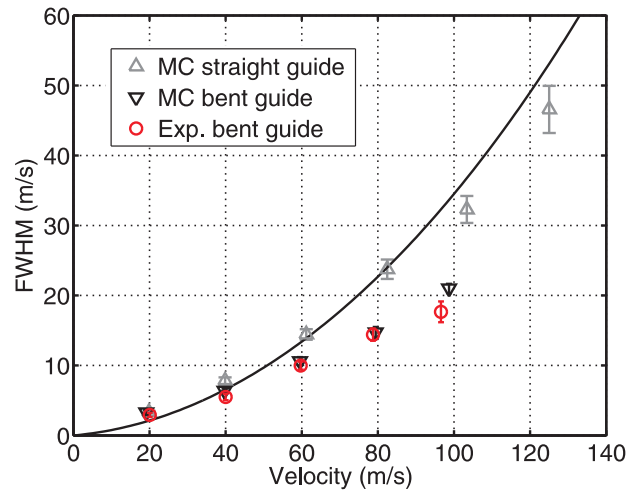}
\caption{(Color online) FWHM of the velocity distributions are presented as a function of the mean velocity of the distribution for both the experimental data (red circles) and the Monte-Carlo simulations for the bent guide (black inverted triangles) and the straight guide (gray triangles). The widths are obtained from fits of the velocity distributions with Voigt profiles. The solid curve represents the analytical model for $\tau=244\,\mu s$ (see text and Eq. $3$). There are no points for the bent guide for velocities above $100\,\rm{m/s}$ since these velocities approach the longitudinal cut-off velocity of our electric guide, and hence the molecular flux is strongly reduced (see Fig. 5).}
\label{fig6}
\end{center}
\end{figure}

To be able to verify the analytical model and to determine $\tau$, we have performed simulations for a segmented straight guide with experimental conditions similar to the ones in the real experiment with the bent guide. The FWHM of the velocity distributions of the molecules reaching the QMS have been fitted to $\Delta v_z$, resulting in $\tau = 244\, \mu s$ as shown in Fig.~\ref{fig6}. This value agrees with our estimate of $\sim\,100\,\mu s$ based on the typical field strength of our deflection field. Note that the mean longitudinal velocities tend to shift to values higher than $v_{z,\,\rm set}$, especially for large $v_{z,\,\rm set}$ (See gray points in Fig. 6). This effect is caused by non-deflected molecules as well, which are more likely to reach the detector at high longitudinal velocities. For $v_{z,\,\rm set} \gtrsim 120\,\rm m/s$ disagreement between the analytical model and the simulation appears due to the limited range of generated velocities. This shows that molecules with velocities beyond the generated range of velocities contribute to these pulses. That is why bends are required to avoid contributions from high-velocity molecules and to keep the mean velocity close to $v_{z,\,\rm set}$.

The FWHM of the velocity distributions for the bent guide are presented as a function of the mean velocity of the distribution for both the experimental data and the Monte-Carlo simulations in Fig.~\ref{fig6} (red circles and inverted triangles, respectively) as well. The results were obtained from fits of the velocity distributions with Voigt profiles. Good agreement is found between experimental data and simulations. This indicates that the main broadening effects, finite overlap times and contribution of residual molecules, are well understood. Compared to the straight guide, the widths increase less strongly at high $v_{z,\,\rm set}$. This is caused by the second bend, which accepts molecules only while being in guiding configuration whereas in a straight guide larger contributions arise from residual molecules (which can survive the deflection fields during time $\tau$). This clearly shows that the second bend reduces the broadening of the peaks.

An interesting aspect is the effect of the different elements of the guide on the behavior of the molecular pulses, which is important for optimization of the experimental setup. Towards this end, we looked at the results from the Monte-Carlo simulations. We recorded the instantaneous velocities of the propagated molecules at the beginning of the guide and at the detector and compared them with the velocities obtained from their time of flight. The pulses in the time domain for the velocities of $80\,\rm{m/s}$ and $100\,\rm{m/s}$ produced at the beginning of the guide are much narrower than the ones produced at the detector. This can be explained by velocity dispersion in the guide. Another reason for this effect, however, might be velocity mixing, which is most likely to occur in the bends of the molecular guide. To study this effect we compared the simulated initial (at the entrance of the guide) velocity distribution of the molecules reaching the detector with their simulated final velocity distribution at the position of the detector. The two velocity distributions are very similar, which implies that the longitudinal velocities do not alter strongly during the propagation in the guide. Therefore, we conclude that no observable velocity mixing occurs in the bends and that the only reason for the pulse broadening in the time domain under our experimental conditions is the velocity dispersion.

\section{Outlook}
\label{summary}

Our method offers prospects for further improvements. As long as the velocity pulses do not overlap, the pulse width depends only on the total length of the guide and the strength of the deflection field. Smaller pulse widths can be obtained using higher deflection fields in non-guiding configuration to eliminate more efficiently contributions of rectilinearly flying molecules. This can be achieved by supplying a higher positive voltage to the non-grounded electrode during off-configuration (see Fig.~\ref{fig4}). From simulations for a straight-guide geometry, we estimate that increasing the voltage from $4\,$kV to $8\,$kV will decrease the relative pulse width from $22\,\%$ (see Fig.~\ref{fig6}) to $14\,\%$ for $v_{z, set}\,=\,60\,$m$/$s. Another approach to minimize these undesired contributions is to use a bent{\color{red}-}guide geometry, in which the molecules are more easily expelled from the guide during the deflection period.
Ultimately the pulse width will depend only on the overlap time used. In our setup the relative contribution amounts to $5\,\%$ of $v_{z,\rm set}$, which can even be further reduced by applying shorter overlap times at the expense of a reduced number of molecules per pulse. Higher overall fluxes can be obtained by running the experiment at a higher repetition rate (up to a few hundred Hertz) without sacrificing on width. By segmenting the guide into more and smaller parts and employing faster switching times, multiple pulses can be stored in the guide at the same time, resulting in even higher rates.

The possibility to tune the width of the velocity independent of the set velocity adds a new degree of freedom to the velocity-filtering technique, which can easily be varied and optimized. In continuous mode the beam is characterized by high density and high flux with a broad velocity spread, whereas pulses with small widths in velocity are obtained at lower density and flux. This technique combined with a buffer-gas cooled source has promising applications in collision and chemistry experiments at low energies. For example, in the collision experiments of \cite{Willitsch2008}, velocity filtered and guided molecules react with cold Coulomb-crystallized ions. Sofar the energy-dependence of the scattering rates has been studied by lowering the voltages on the guide electrodes resulting in a lower cutoff velocity $v_{z,{\rm max}}$ and therefore an on average lower $v_z$. Unfortunately, lowering the guiding field also reduces $v_{t,{\rm max}}$.
This can be avoided by using a segmented guide, which has also the advantage to narrow the velocity around a set velocity.
In addition, the rotational temperature can be independently adjusted by varying the density of the buffer gas. Cold collisions can also be studied by directing velocity-selected pulses into a dense beam as has been shown in \cite{Gilijamse2006} or by guiding the slow molecules into a long-lived trapped sample. Velocity-selected molecules could then collide with laser-cooled atoms stored in an electric trap \cite{Rieger2007,Schlunk2007}.
Another realm of application of our method could be molecular interferometry with guidable molecules. A major challenge in this field is the production of molecules with well-defined and controllable velocities \cite{Brezger2003,Gerlich2007}. Pulses from a segmented guide could be tuned to a setting at which highest contrast is obtained.  Therefore, the segmented-guiding technique could be well-suited for such experiments with molecules having sufficiently large dipole moments.

\section{Acknowledgements}
\label{Acknowledgements}
Financial support from Deutsche Forschungsgemeinschaft (Cavity-Mediated Molecular Cooling CMMC and Munich-Centre for Advanced Photonics MAP) is gratefully acknowledged. We appreciate initial discussions on the topic with Sadiq A. Rangwala.

\clearpage

\end{document}